\documentstyle[11pt,amssymb]{article}

\makeatletter
\@addtoreset{equation}{section}
\makeatother

\def\ben{\begin{equation}}
\def\een{\end{equation}}

\let\a=\alpha    
    
  \let\n=\nu

\let\C=\Chi

\def\nn{\nonumber} \def\bd{\begin{document}} \def\ed{\end{document}}
\def\ds{\documentstyle} \let\fr=\frac \let\bl=\bigl \let\br=\bigr
\let\Br=\Bigr \let\Bl=\Bigl
\let\bm=\bibitem
\let\na=\nabla
\let\pa=\partial \let\ov=\overline
\newcommand{\be}{\begin{equation}}
\newcommand{\ee}{\end{equation}}
\def\ba{\begin{array}}
\def\ea{\end{array}}
\def\ft#1#2{{\textstyle{{\scriptstyle #1}\over {\scriptstyle #2}}}}
\def\fft#1#2{{#1 \over #2}}
\def\del{\partial}
\def\vp{\varphi}
\def\sst#1{{\scriptscriptstyle #1}}
\def\oneone{\rlap 1\mkern4mu{\rm l}}
\def\td{\tilde}
\def\wtd{\widetilde}
\def\ie{{\it i.e.\ }}
\def\dalemb#1#2{{\vbox{\hrule height .#2pt
        \hbox{\vrule width.#2pt height#1pt \kern#1pt
                \vrule width.#2pt}
        \hrule height.#2pt}}}
\def\square{\mathord{\dalemb{6.8}{7}\hbox{\hskip1pt}}}
\newcommand{\ho}[1]{$\, ^{#1}$}
\newcommand{\hoch}[1]{$\, ^{#1}$}
\newcommand{\bea}{\begin{eqnarray}}
\newcommand{\eea}{\end{eqnarray}}
\newcommand{\ra}{\rightarrow}
\newcommand{\lra}{\longrightarrow}
\newcommand{\Lra}{\Leftrightarrow}
\newcommand{\ap}{\alpha^\prime}
\newcommand{\bp}{\tilde \beta^\prime}
\newcommand{\tr}{{\rm tr} }
\newcommand{\Tr}{{\rm Tr} }
\def\0{{\sst{(0)}}}
\def\1{{\sst{(1)}}}
\def\2{{\sst{(2)}}}
\def\3{{\sst{(3)}}}
\def\4{{\sst{(4)}}}
\def\5{{\sst{(5)}}}
\def\6{{\sst{(6)}}}
\def\7{{\sst{(7)}}}
\def\8{{\sst{(8)}}}
\def\n{{\sst{(n)}}}
\def\cA{{{\cal A}}}
\def\cF{{{\cal F}}}
\def\tV{\widetilde V}
\def\tW{\widetilde W}
\def\tH{\widetilde H}
\def\tE{\widetilde E}
\def\tF{\widetilde F}
\def\tA{\widetilde A}
\def\im{{{\rm i}}}
\def\tY{{{\wtd Y}}}
\def\ep{{\epsilon}}
\def\vep{{\varepsilon}}
\def\bD{{{\bar D}}}
\def\alp{{{\a'}^3}}
\def\bD{{{\bar D}}}
\def\R{{{\Bbb R}}}
\def\C{{{\Bbb C}}}
\def\H{{{\Bbb H}}}
\def\CP{{{\Bbb C}{\Bbb P}}}
\def\RP{{{\Bbb R}{\Bbb P}}}
\def\Z{{{\Bbb Z}}}
\def\bA{{{\Bbb A}}}
\def\bB{{{\Bbb B}}}
\def\bC{{{\Bbb C}}}
\def\bR{{{\Bbb R}}}
\def\bD{{{\Bbb D}}}
\def\bE{{{\Bbb E}}}
\def\bZ{{{\Bbb Z}}}
\def\Re{{{\frak{Re}}}}
\def\Im{{{\frak{Im}}}}
\def\cosec{{\,\hbox{cosec}\,}}
\def\Gm{{\Gamma_{\!\! -}}}
\def\Gp{{\Gamma_{\!\! +}}}
\def\stan{{standard }}
\def\nonstan{{supernumerary }}

\def\cosech{{\hbox{cosech}}}

\def\etcyc{{\hbox{and cyclic}}}
\def\btheta{{\bar\theta}}

\newcommand{\tamphys}{\it Center for Theoretical Physics,
Texas A\&M University, College Station, TX 77843, USA}

\newcommand{\mitchell}{\it George P. \& Cynthia W.
Mitchell Institute for Fundamental Physics,\\
Texas A\&M University, College Station, TX 77843-4242, USA}
\newcommand{\umich}{\it Michigan Center for Theoretical Physics,
University of Michigan\\ Ann Arbor, MI 48109, USA}
\newcommand{\upenn}{\it Department of Physics and Astronomy,
University of Pennsylvania, Philadelphia,  PA 19104, USA}
\newcommand{\SISSA}{\it  SISSA-ISAS and INFN, Sezione di Trieste\\
Via Beirut 2-4, I-34013, Trieste, Italy}

\newcommand{\newton}{\it Isaac Newton Institute for Mathematical Sciences,\\
20 Clarkson Road,  University of Cambridge,
Cambridge CB3 0EH, UK}

\newcommand{\ihp}{\it Institut Henri Poincar\'e\\
  11 rue Pierre et Marie Curie, F 75231 Paris Cedex 05}

\newcommand{\damtp}{\it DAMTP, Centre for Mathematical Sciences,
 Cambridge University\\  Wilberforce Road, Cambridge CB3 OWA, UK}
\newcommand{\itp}{\it Institute for Theoretical Physics, University of
California\\ Santa Barbara, CA 93106, USA}

\newcommand{\imperial}{\it The Blackett Laboratory, Imperial College London\\
Prince Consort Road, London SW7 2AZ. }

\newcommand{\auth}{
H. L\"u\hoch{\ddagger1},
C.N. Pope\hoch{\ddagger1} and K.S. Stelle\hoch{\star2} }

\begin{document}
\begin{flushright}
\hfill{
MIFP-05-20 \ \
Imperial/TP/050901}\\
\hfill{
\bf hep-th/0509057}
\end{flushright} 

\begin{center}  

{\Large {\bf Generalised Holonomy for Higher-Order Corrections to 
 Supersymmetric Backgrounds in String and M-Theory
}}   

\vspace{15pt}

\auth

\vspace{7pt}
{\hoch{\ddagger}\mitchell}

\vspace{7pt}
{\hoch{\star}\imperial} 

\vspace{30pt}

\underline{ABSTRACT}
\end{center}  

   The notion of {\it generalised structure groups} and {\it generalised 
holonomy groups} has been introduced in supergravity, in order to discuss
the spinor rotations generated by commutators of supercovariant derivatives
when non-vanishing form fields are included, with their associated 
gamma-matrix structures that go beyond the usual $\Gamma_{MN}$ of 
the Riemannian connection.  In this paper we investigate the generalisations
to the usual Riemannian structure and holonomy groups that result from the
inclusion of higher-order string or M-theory corrections in the 
supercovariant derivative.  Even in the absence of background form fields, 
these corrections introduce additional terms $\Gamma_{M_1\cdots M_6}$ in
the supercovariant connection, and hence they lead to enlarged structure and
holonomy groups. In some cases, the corrected equations of motion force
form fields to become non-zero too, which can further enlarge the groups.
Our investigation focuses on the generalised structure and holonomy groups in
the transverse spaces $K_n$ of (Minkowski)$\times K_n$ backgrounds for
$n=6$, 7, 8 and 10, and shows how the generalised holonomies allow the
continued existence of supersymmetric backgrounds even though the usual
Riemannian special holonomy is destroyed by the inclusion of the string or
M-theory corrections.

{\vfill\leftline{}\vfill
\vskip 10pt \footnoterule
{\footnotesize \hoch{1}
Research supported in part by DOE grant DE-FG03-95ER40917
\vskip -12pt} \vskip 14pt
{\footnotesize \hoch{2}
Research supported in part by the EC under MRTN
contract MRTN-CT-2004-005104 and by PPARC \\
$\phantom{xxxx}$ under rolling grant \uppercase{PPA/G/O}/2002/00474, PP/D50744X/1
\vskip -12pt}  \vskip  14pt
}

\pagebreak
\setcounter{page}{1}

\tableofcontents
\addtocontents{toc}{\protect\setcounter{tocdepth}{2}}
\newpage 

\section{Introduction}

   Corrections at order $\alp$ in tree-level string theory imply that
leading-order supersymmetric backgrounds, such as the product of
four-dimensional Minkowski spacetime and a Ricci-flat Calabi-Yau
6-manifold $K_6$, are modified so that the internal space is no longer
Ricci flat \cite{grisaru,fpss,freemanpope,grosswitten}.  
Although this means that the $SU(3)$ holonomy group that
ensured the existence of a covariantly-constant spinor in $K_6$ is
enlarged (to $U(3)$ in this case), it is expected that
supersymmetry will survive in the deformed background, because of
compensating correction terms in the supersymmetry transformation
rules.  No direct and complete calculation of the order $\alp$
corrections to the supersymmetry transformation rules exists, and so a
fully explicit proof of the preservation of supersymmetry of the
deformed backgrounds has not been possible.  However, in \cite{cfpss} it was
noted that the supersymmetry of deformed Calabi-Yau 6-manifold backgrounds
would persist if a certain correction term were added to the usual
covariant derivative appearing in the gravitino transformation rule.
Specifically, the proposal in \cite{cfpss} was to
modify the usual covariant derivative
\be
\nabla = d + \ft14 \omega_{AB}\,\Gamma^{AB}
\ee
to $\widetilde\nabla =\nabla + Q$, with components given by
\be
\widetilde\nabla_M \equiv \nabla_M + Q_M= \nabla_M - {3c\over4} \alp \,
\left[ (\nabla^N\, R_{MP N_1 N_2})\, R_{NQ N_3 N_4}\,
R^{PQ}{}_{N_5 N_6}\right]\, \Gamma^{N_1 \cdots N_6} +
{\cal O}\left(\alpha'{}^4\right)\,,\label{covmod}
\ee
where $c$ is a certain purely numerical constant.
To order $\alp$, this has the integrability condition
\be
[\wtd\nabla_M,\wtd\nabla_N]\eta = \ft14 R_{MNPQ}\, \Gamma^{PQ}\eta  + 
           2 \nabla_{[M} Q_{N]}\eta=0\,.
\label{int1}
\ee
It was shown from this that the existence of a Killing spinor $\eta$
satisfying $\wtd\nabla_M\, \eta=0$ requires a non-vanishing Ricci 
tensor that is
precisely the one implied by the known order $\alp$ corrections to
Einstein's equations in string theory \cite{cfpss}.

   More recently, the correction term in (\ref{covmod}) has been
probed in more detail, by considering the deformations of
supersymmetric backgrounds (Minkowski)$\times K_n$ in string and
M-theory for a variety of further special-holonomy manifolds $K_n$,
with dimensions $n=7$ \cite{g2paper}, and $n=8$ and $n=10$ 
\cite{spin7paper}.  In the case of M-theory
backgrounds, the correction terms in question are very similar in form
to the tree-level $\alp$ corrections in string theory, but associated
now with one-loop string corrections.  Specifically, it has been shown
that for deformations of seven-dimensional $G_2$ holonomy spaces,
eight-dimensional Spin(7) holonomy spaces, and ten-dimensional $SU(5)$
holonomy spaces, the correction (\ref{covmod}) that was originally proposed
purely for 6-dimensional Calabi-Yau backgrounds is in fact sufficient also
to imply the preservation of supersymmetry for all the high-dimensional
special-holonomy spaces too \cite{g2paper,spin7paper}.

   In view of the apparent universality of the correction term in
(\ref{covmod}), and in the absence of any other direct calculation of the
corrections to the supersymmetry transformation rules, it is of interest
to pursue further the implications that follow from supposing that
(\ref{covmod}) might in fact be the only relevant correction term.
In this paper, we shall study the notion of {\it generalised holonomy}
\cite{dust,duli1,hull}.  However, unlike the earlier work, our focus
will be on the modifications to generalised holonomy associated
with the inclusion of higher-order correction terms in string and M-theory.

   If we consider a purely gravitational background, with vanishing
field strengths, then from (\ref{int1}) we see that the generalised
structure group is generated by the Dirac matrices $\Gamma_{MN}$ and
$\Gamma_{N_1\cdots N_6}$.  Additionally, one must include any further
Dirac matrices that result from taking commutators of these, until
closure is achieved.  In $D=11$ the 55 matrices $\Gamma_{MN}$ generate
$SO(1,10)$, and these are supplemented by the 462 matrices
$\Gamma_{N_1\cdots N_6}$.  The commutators $[\Gamma_{M_1\cdots
M_6},\Gamma_{N_1\cdots N_6}]$ generate 10-Gamma terms, which are dual
to the $\Gamma_M$ themselves, together with 6-Gamma and 2-Gamma terms
which are already included.  Thus in total we obtain closure on the
$55+462+11=528$ generators $\{\Gamma_{MN}, \Gamma_{M_1\cdots M_6},
\Gamma_M\}$.  Writing $M=(0,i)$, it can be seen that these divide into 
Hermitean and anti-Hermitean generators as follows:
\bea
\hbox{Hermitean}: && \Gamma_i \,,\quad \Gamma_{0i}\,,\quad 
  \Gamma_{0i_1\cdots i_5}\,,\qquad\qquad 10+10+252= 272\,,\nn\\
\hbox{anti-Hermitean}:&& \Gamma_0\,,\quad \Gamma_{ij}\,,\quad 
   \Gamma_{i_1\cdots i_6}\,,\qquad\qquad 1+45+210= 256\,.
\eea
Thus we have 256 compact generators and 272 non-compact, giving rise to 
the algebra  $Sp(32)$.  This is maximally split, with 256 compact and
256+16 non-compact generators, where the 16 form the Cartan subalgebra.
It should be
emphasised that this generalised structure group is specific to $D=11$
backgrounds where the 4-form field strength vanishes; it is the
generalisation of $SO(1,10)$ that arises due to higher-order
corrections in those cases where the 4-form does not play a r\^ole.

    If one does include the 4-form as well, then in fact the generalised
structure group is the same whether or not the higher-order corrections
in (\ref{covmod}) are included.  To see this, consider first the standard
leading-order terms in the supercovariant derivative:
\be
D_M= \nabla_M - \ft1{288}\, F_{N_1\cdots N_4}\, \Gamma_M{}^{N_1\cdots N_4} 
   + \ft1{36}\, F_{MN_1\cdots N_3}\, \Gamma^{N_1\cdots N_3}\,.
\label{supercov}
\ee
Here, the commutator of supercovariant derivatives gives the $\Gamma_{MN}$,
$\Gamma_{M_1\cdots M_3}$ and $\Gamma_{M_1\cdots M_5}$ matrices.  Denoting 
$\Gamma_{M_1\cdots M_p}$ collectively by $\Gamma_{(p)}$, we therefore have
$\Gamma_{(2)}$, $\Gamma_{(3)}$ and $\Gamma_{(5)}$.  Further commutators of
these will generate in addition $\Gamma_{(1)}$ and $\Gamma_{(4)}$.
The total numbers of $\{\Gamma_{(1)}, \Gamma_{(2)}, 
\Gamma_{(3)}, \Gamma_{(4)}, \Gamma_{(5)}\}$ give $11+55+165+330+462=1023$
generators.  These split into Hermitean and anti-Hermitean generators
as follows:
\bea
\hbox{Hermitean}: && \Gamma_i \,,\ \ \Gamma_{0i}\,,\ \ 
  \Gamma_{0i_1\cdots i_5}\,,\ \ \Gamma_{0ij}\,,\ \ 
       \Gamma_{ijk\ell}\,,\quad 10+10+252 +45+210= 527\,,\nn\\
\hbox{anti-Hermitean}:&& \Gamma_0\,,\ \ \Gamma_{ij}\,,\ \ 
   \Gamma_{i_1\cdots i_6}\,,\ \ \Gamma_{ijk}\,,\ \ 
    \Gamma_{0ijk}\,,\quad 1+45+210+120+120= 496\,.\nn
\eea
In this case we have 496 compact and 527 non-compact generators,
closing on the
algebra $SL(32,\R)$. This is again is maximally split, since we have 
496 compact and 496+31 non-compact generators, where the 31 form the 
Cartan subalgebra.  Thus we have reproduced the result in \cite{hull}, 
that $SL(32,\R)$ is the generalised structure 
group.  Inclusion of the higher-order corrections (\ref{covmod})
merely adds further $\Gamma_{(6)}$ terms, which are equivalent to 
the $\Gamma_{(5)}$ that is already present at leading order.

   In a similar fashion, we may consider the situation in $D=10$ in
string theory.  Again, if we first consider the situation of purely
gravitational backgrounds, with all form fields vanishing, 
then we find that closure is achieved on
the set of $45+210=25$ matrices  $\{ \Gamma_{(2)}, \Gamma_{(6)}\}$.   
These generate the algebra $SL(16,\R)$.  Again, if we consider the 
general case of backgrounds with non-vanishing form fields too, then the
inclusion of the higher-order correction term (\ref{covmod}) adds 
no new Dirac matrix structures, and so one has the same generalised
structure group $SL(32,\R)$ as at leading order.

   The layout of this paper is as follows.  In section 2, we introduce the
notion of the ``Generalised Transverse Structure Group,'' which is defined
by viewing the background for a deformed vacuum configuration 
from the viewpoint of the space transverse to the lower-dimensional 
Minkowski spacetime.  Effectively, this amounts to performing a Kaluza-Klein 
dimensional reduction on the Minkowski spacetime, allowing us to factor
this out and discuss the generalised structure group (and ultimately, the
generalised holonomy group) purely from the viewpoint of the $n$-dimensional 
transverse space $K_n$.  By this means, we can focus on the generalisation 
of the standard $SO(n)$ transverse structure group, and its special-holonomy
subgroup.  Of course the Kaluza-Klein reduction amounts to just a trivial 
direct-product factorisation in cases where the leading-order 
(Minkowski)$\times K_n$ direct-product structure is preserved by the
string or M-theory corrections.  However, in the case of backgrounds with
an eight or ten-dimensional transverse space, the original direct-product
structure is deformed to a warped product by the string or M-theory 
corrections, and then the Kaluza-Klein reduction on the Minkowski factor
becomes non-trivial.

   In section 3, we consider the generalised transverse holonomy group
for supersymmetric backgrounds of the form (Minkowski)$\times K_7$,
where $K_7$ is a space of $G_2$ holonomy at leading order.  By
considering a specific example of a $G_2$ holonomy space, we show,
using some results from \cite{g2paper}, how the deformed solution where
string or M-theory corrections are included continues to be supersymmetric,
and we determine the associated generalised transverse holonomy group.  
One can see from this result, and the embedding of the generalised 
transverse holonomy group in the generalised transverse structure group,
how it is that there is still a singlet in the decomposition, and hence 
still a surviving supersymmetry, even though the deformation leads to
an enlargement of the standard transverse holonomy from $G_2$ to 
$SO(7)$.  

   In sections 4 and 5, we carry out analogous investigations of the
generalised transverse holonomy groups for supersymmetric backgrounds
of the form (Minkowski)$\times K_n$ for $n=8$ and $n=10$.  In each
case, we consider an example space $K_n$ that initially has special
holonomy, namely Spin(7) for $n=8$, and $SU(5)$ for $n=10$.  Following
the discussions in \cite{g2paper} and \cite{spin7paper}, we show how
the supersymmetry is preserved when string or M-theory corrections are
included, and how the generalised transverse holonomy groups in each
case are consistent with the continued existence of a singlet in the
decomposition of the generalised transverse structure groups.
Finally, in section 6, we end with conclusions, including a summary of
our results on the generalised transverse structure and holonomy
groups in the various dimensions.

\section{Generalised Structure Groups in the Transverse Space}

   When considering the holonomy groups of supersymmetric
compactifications in M-theory or string theory, a full treatment
involves addressing this question in the entire $D=11$ or $D=10$
spacetime.  It is often convenient, however, to factor the discussion
into a consideration of the holonomy, or its non-Riemannian
generalisation, purely within the transverse space orthogonal to the
lower-dimensional spacetime.  At leading order, prior to the
introduction of the higher-order string or M-theory corrections, this
is rather clear, since the leading-order supersymmetric backgrounds
under discussion are all of the form (Minkowski)$_d\times K_n$, where
$K_n$ is a Ricci-flat $n$-manifold of special holonomy, and $d+n=10$
or 11.  Once the higher-order corrections are taken into account, the
original direct-product background may, in some cases, be deformed into
a warped product.  However, it turns out that we can still discuss the
notion of a generalised holonomy group within the $n$-dimensional
transverse space.  Before doing this, we must first consider the
generalised structure group in the transverse space, which characterises
the ``generic'' holonomy of a non-supersymmetric vacuum.

    To be more precise, we shall introduce the notion first of a
``Generalised Transverse Structure Group.''  The idea here is the
following.  The question of principle interest to us in this paper is
how the generalised holonomy of deformed supersymmetric backgrounds
differs from their ordinary holonomy at leading order. Additionally,
we are interested in seeing how the generalised holonomy of a deformed
supersymmetric background differs from the generalised holonomy of a
deformed non-supersymmetric background.  The viewpoint that we shall
take in all of our discussion is that, wherever the field equations
allow, we shall consider purely gravitational backgrounds.  If the
field equations force the original purely-gravitational background to
acquire non-vanishing form-field contributions at order $\alp$ in a
perturbative expansion scheme, then we shall (as we must) include such
contributions.  However, the only form-field terms we shall consider
are those of this kind, which are forced by the purely gravitational
source-terms. One reason for focussing on such cases is that little is
known about the correction terms at order $\alp$ involving the form
fields in string theory,\footnote{Partial results up to terms of the
form $(DF)^2 R^2$ and $(DF)^4$ have been obtained recently in
\cite{peplst}.} whereas the gravitational parts at this order are
fully known.  As long as one studies in perturbation theory up to
order $\alp$, starting from purely gravitational backgrounds, then the
lack of knowledge about the form fields is not a problem.  However,
one cannot start with backgrounds where the form fields are ``large,''
and deform these by order $\alp$ corrections, given our present
incomplete knowledge of the form-field correction terms. Of course, one
could allow the form fields to be turned on independently at order $\alp$, in
addition to whatever order $\alp$ contributions are forced by the
deformation of a purely gravitational background, but this would seem
to be somewhat artificial, introducing complication but little extra
insight.

    At leading order, for example, we may consider
(Minkowski)$_d\times K_n$, where $d+n=10$ or 11, and $K_n$ is
Ricci-flat.  By choosing $K_n$ to have special holonomy rather than
generic $SO(n)$ holonomy one gets supersymmetric compactifications.
One may likewise consider the $\alp$-induced deformations of the
leading-order (Minkowski)$_d\times K_n$ backgrounds, and again make
the comparison, now at the level of generalised holonomy, between the
cases with and without supersymmetry.

   Having made the restriction to generalised structure groups for
backgrounds which are deformations of purely gravitational
leading-order backgrounds, it is then natural to go one step further,
and focus where possible on the structure group in the $n$-dimensional
transverse space of the manifold $K_n$.  The situation is most
straightforward for $n=6$ and $n=7$, since then the deformed
background retains the direct-product form (Minkowski)$_d\times K_n$,
and only the geometry of $K_n$ itself is deformed as a result of the
inclusion of the string or M-theory corrections.  For $n=8$, there are
two cases to consider, depending upon whether on the one hand one is
considering tree-level string corrections, in which case the
direct-product form (Minkowski)$_d\times K_8$ is maintained, or
whether on the other hand one is considering one-loop string or
M-theory corrections, in which case the equations force a
warped-product structure and a non-vanishing form-field.  For the
$n=10$ case, namely (Minkowski)$_1\times K_{10}$ in M-theory, a warped
factorisation again occurs.

   In all the above cases, whether warped or not, the
eleven-dimensional metric describing the deformed solution takes the
form
\be
d\hat s_{11}^2 = e^{2\alpha\varphi}\, ds_n^2 + e^{2\beta\varphi}\, 
   dx^\mu dx_\mu\,,
\ee
where the constants $\alpha$ and $\beta$ are related by $\beta=
-\alpha(n-2)/(11-n)$, and, in the case of warped products, $\varphi$
is a function of the coordinates on the transverse space whose metric
is $ds_n^2$.  One can think of performing a dimensional reduction on
the flat $(11-n)$-dimensional Minkowski spacetime, in which case
$ds_n^2$ acquires an interpretation as the Einstein-frame metric on
the lower-dimensional transverse space.  Similarly, if the form field is
non-zero (in the sense we discussed above), then we perform a
standard type of Kaluza-Klein reduction on this too.

   Let us first consider a generic background of this type, where we
do not yet assume that there is any supersymmetry. The {\it 
generalised transverse
structure group} is generated by the closure of the set of gamma
matrices appearing in the dimensional reduction of the supercovariant
derivative (\ref{supercov}), with the corrections (\ref{covmod}) to
the covariant derivative included.  The {\it generalised transverse
holonomy group} is then defined analogously, for the specific background
solution under consideration.  If this background is supersymmetric,
its generalised transverse holonomy group will be a proper subgroup
of the generalised transverse structure group.

   In the following subsections, we shall consider the
generalised transverse structure groups for the various
dimensions of interest.  In particular,
for transverse spaces $K_n$ with dimension $n\le 7$, the entire
discussion can be given in the absence of form fields.

\subsection{Generalised transverse 
 structure group for $n=6$}\label{n6structuresec}

   We are considering here situations where the undeformed background is
purely gravitational, with no form fields turned on.  In cases where
the curvature in this background is confined to six dimensions, the
$\alp$-corrected equations of motion do not lead to any non-vanishing
sources for the form fields, in this perturbative discussion, and so
the background will remain purely gravitational.  From (\ref{covmod}) and
(\ref{int1}), we see that the commutators of supercovariant
derivatives will yield terms involving $\Gamma_{(2)}$ and
$\Gamma_{(6)}$ in the six dimensions, where we again use the notation
$\Gamma_{(p)}$ to denote the set of all $p$-index antisymmetrised
Dirac matrices $\Gamma_{i_1\cdots i_p}$.  In six dimensions $\Gamma_{(6)}$
is simply proportional to $\im \Gamma_7$, where $\Gamma_7$ is the chirality
operator in $D=6$ (squaring to $+1$).  It is then easy to see that the set
of matrices $\Gamma_{(2)}$ and $\Gamma_7$ close under commutation, giving
the generalised transverse structure group $SO(6)\times U(1)$.

\subsection{Generalised transverse structure 
group for $n=7$}\label{n7structuresec}

    In this case, since the form fields can again remain zero when the
higher-order string theory or M-theory corrections are taken into account, it
again suffices to consider just the gravitationally-corrected
covariant derivative appearing in (\ref{covmod}).  In a generic curved
background, the structure group associated with (\ref{covmod}) will be
generated by the Dirac matrices $\Gamma_{(2)}$ and $\Gamma_{(6)}$,
together with any additional matrices required by closure.  In $n=7$
dimensions $\Gamma_{(6)}$ is dual to $\Gamma_{(1)}$, and so we
immediately achieve closure on the matrices $\Gamma_{(2)}$ and
$\Gamma_{(1)}$.  These generate the group $SO(8)$.  Thus we conclude
that for the class of purely gravitational seven-dimensional
backgrounds that we are considering here, the standard $SO(7)$
structure group (the tangent-space group) is enlarged to the
generalised transverse structure group $SO(8)$.

\subsection{Generalised transverse structure group for 
             $n=8$}\label{n8structuresec}

   The discussion of the generalised transverse structure group 
in $K_8$ for backgrounds of the form 
(Minkowski)$\times K_8$ at leading order takes a sightly
different form depending on whether one is considering the effect of
tree-level corrections in string theory, or alternatively 1-loop
string corrections or (equivalently) M-theory corrections.  In the case
of tree-level corrections in string theory, an initial purely gravitational
background (Minkowski)$_2\times K_8$ remains purely gravitational, in the
sense that the corrected equations of motion do not require that form
fields become non-zero.

    If, on the other hand, we consider the effect of 1-loop string
corrections or M-theory corrections, the equations of motion now imply
that form fields must also become non-vanishing, and furthermore that the
ten-dimensional or eleven-dimensional metric deforms into a warped
product.  The discussion of the one-loop string and the M-theory cases
is essentially identical, and so when we consider this situation we
shall, for definiteness, discuss it in the context of M-theory.

   Let us first, however, consider the situation at tree level.  We
need only consider the generalised structure group for the modified
covariant derivative given in (\ref{covmod}).  The $\Gamma_{(6)}$
terms in (\ref{covmod}) can be dualised to $\Gamma_9\, \Gamma_{(2)}$,
and thus we achieve closure on the matrices $\Gamma_{ij}$ and
$\Gamma_9\, \Gamma_{ij}$.  If we define
\be
X^+_{ij} = \ft12(1+\Gamma_9)\, \Gamma_{ij}\,,\qquad
X^-_{ij} = \ft12(1-\Gamma_9)\, \Gamma_{ij}\,,
\ee
it is immediately apparent that the $X^+_{ij}$ and $X^-_{ij}$ matrices
generate two commuting copies of $SO(8)$, which we shall denote by
$SO(8)_+$ and $SO(8)_-$ respectively.  Thus the generalised transverse 
structure group for purely gravitational 8-dimensional transverse spaces is
$SO(8)_+\times SO(8)_-$.

   Turning now to M-theory, it is known that in addition to the
corrections to the Einstein equation for the internal space $K_8$, the
corrections to the form-field equation imply that $F_\4$ must become
non-zero \cite{hawtay,becbec,becbec2,brcvna,spin7paper,konst}. 
Specifically, in a perturbative analysis of the deformation
of the leading-order solution, one finds that the eleven-dimensional
metric warps, assuming the form
\be
d\hat s_{11}^2 = e^{2A}\, dx^\mu\, dx_\mu + e^{-A}\, ds_8^2\,,
\ee
whilst the 4-form is given by 
\be
\hat F_\4 = d^3 x\wedge df\,.
\ee
The warp factor $A$ \cite{becbec2,spin7paper,konst} and the scalar $f$ satisfy
\be
\square A= \fft{\beta}{1728}\, Y_2\,,\qquad
\square f =m \beta\, (2\pi)^4\, {* X_\8}\,,
\ee
where $Y_2$ is a certain multiple of the eight-dimensional Euler
integrand,
\be
Y_2 = \ft{315}{2} \hat R^{[M_1M_2}{}_{M_1 M_2}\, \cdots
    \hat R^{M_7 M_8]}{}_{M_7 M_8}\,.\label{y2exp}
\ee
and the 8-form $X_\8$ is given by
\be
X_\8 = \fft1{192\, (2\pi)^4}\, [
 \tr\, \Theta^4 - \ft14 (\tr\, \Theta^2)^2]\,,
\ee
For this general class of warped backgrounds, the 
supercovariant derivative
in the eight transverse dimensions takes the form \cite{spin7paper}
\be
\hat D_i = \oneone\otimes \nabla_i - \ft14 \del_j A\, \oneone\times
\Gamma_i{}^j -\ft1{12}\del_j f\, \oneone\otimes \Gamma_i{}^j\, \Gamma_9
   +\ft16 \del_i f\, \oneone\otimes \Gamma_9 + \oneone\otimes Q_i\,,
\label{did8}
\ee
where $\Gamma_i$ denotes the Dirac matrices in the eight-dimensional
transverse space, and $\Gamma_9$ is the associated chirality operator.
 From (\ref{did8}), one can show \cite{spin7paper} that the configurations
discussed in \cite{hawtay,becbec,brcvna,spin7paper,konst} are supersymmetric
at the leading order in the expansion parameter $\beta$.
We also see that the algebra generated by their commutators again closes
on the generators $\Gamma_{(2)}$ and $\Gamma_9\, \Gamma_{(2)}$, and
so again we conclude that the generalised transverse structure group 
is $SO(8)_+ \times SO(8)_-$.

\subsection{Generalised transverse structure group for $n=10$}

   Configurations with a ten-dimensional curved transverse space can
be discussed only in M-theory.  Again, it has been shown that a
solution that is simply a direct product (Minkowski)$_1\times K_{10}$
at leading order becomes deformed into a warped product, with a
non-vanishing 4-form field, when the M-theory corrections are taken
into account \cite{spin7paper}.  In this case, therefore, we must
necessarily include the effect of the 4-form in the discussion.

   The form of the warped-product metric is \cite{spin7paper}
\be
d\hat s_{11}^2 = -e^{2A}\, dt^2 + e^{-\ft14 A}\, ds_{10}^2\,,\label{11to10}
\ee
whilst the 4-form is given by
\be
\hat F_\4 = G_\3 \wedge dt\,.
\ee
In the ten-dimensional internal space the supercovariant derivative is
then given by \cite{spin7paper}
\be
\hat D_i = \nabla_i-\ft1{16}\nabla_jA\, \Gamma^{ij}
   +\ft{\im}{72} G_{jk\ell} \, \Gamma_i{}^{jk\ell}\, \Gamma_{11}
  - \ft{\im}{12} G_{ijk}\, \Gamma^{jk}\, \Gamma_{11}
   +Q_i\,,
\ee
when expressed in terms of the ten-dimensional $SO(10)$ Dirac matrices
$\Gamma_i$, and the ten-dimensional chirality operator $\Gamma_{11}$.
It therefore follows that the commutators of supercovariant
derivatives close on the matrices
\be
\{ \Gamma_{(2)}, \Gamma_{(4)},
   \im \,\Gamma_{11}\, \Gamma_{(2)}, 
          \im\, \Gamma_{11}\, \Gamma_{(4)} \}\label{24gam}
\ee
in the transverse ten-dimensional space.  These divide into Hermitean and
anti-Hermitean generators as follows:
\bea
\hbox{Hermitean}:&& \im\, \Gamma_{ij}\, \Gamma_{11}\,,\qquad 
     \Gamma_{ijk\ell}\,,\qquad\qquad 45+210= 255\,,\nn\\
\hbox{anti-Hermitean}:&& \Gamma_{ij}  \,,\qquad 
     \im\, \Gamma_{ijk\ell}\, \Gamma_{11}\,,
                 \qquad\qquad 45+210= 255\,,\label{sl16c}
\eea
and so we have 255 compact and 255 non-compact generators.  The 
255 compact generators $\Gamma_{ij}$ and  
$\im\, \Gamma_{ijk\ell}\, \Gamma_{11}$ close on $SU(16)$, and the complete 
set of generators form the algebra $SL(16,\C)$.  Thus the
generalised transverse structure group for deformations of 
original (Minkowski)$_1\times K_{10}$ backgrounds is
$SL(16,\C)$.

   It is worth emphasising that the situation in this case of
(Minkowski)$_1\times K_{10}$ backgrounds in M-theory is quite
different from the (Minkowski)$\times K_n$ for $6\le n\le 8$ examples
that we discussed previously, in that here we are finding a non-compact 
generalised structure group, even though we are focusing purely on the
Euclidean-signature ten-dimensional transverse space.  

   In the following sections, we shall study the generalised
transverse holonomy groups that arise when considering string or
M-theory backgrounds of the form (Minkowski)$\times K_n$, where at
leading order $K_n$ is a special-holonomy space of dimension $n=7$, 8
or 10.  Before moving on to considering these cases, we note that the
case of (Minkowski)$\times K_6$ backgrounds was effectively already
been discussed in the earlier literature \cite{cfpss}.  In this case,
the effect of including the higher-order corrections is to introduce 
$\Gamma_{(6)}$ terms that are dual to $\im\, \Gamma_7$ in $K_6$, and so
the leading-order $SU(3)$ special holonomy of the Calabi-Yau background is
enlarged to $SU(3)\times U(1)$.  This extra $U(1)$ factor generated by
$\im\, \Gamma_7$ is cancelled
by a $U(1)$ contribution proportional to $J^{ij}\, \Gamma_{ij}$ coming
from the spin connection of the deformed background when the supercovariant 
derivative acts on a Killing spinor, thus implying that
supersymmetry is preserved.  As we shall see in the following
sections, the cases of (Minkowski)$\times K_n$ backgrounds with $n>6$
lead to less trivial enlargements of the holonomy groups. 

\section{Generalisation of $G_2$ Transverse Holonomy}

   In section \ref{n7structuresec}, we showed that the generalised
transverse structure group for purely gravitational seven-dimensional
backgrounds is $SO(8)$, which represents an enhancement of the
standard $SO(7)$ Riemannian structure group.  In this section we shall
examine the generalised transverse holonomy group for supersymmetric
solutions with such a seven-dimensional curved transverse space.
Specifically, we shall be interested in the supersymmetric backgrounds
of the form (Minkowski)$_4\times K_7$ at leading order, where $K_7$ is
a Ricci-flat metric of $G_2$ holonomy, and the deformations of these
backgrounds in the face of higher-order string or M-theory
corrections.  As was shown in \cite{g2paper}, the deformed backgrounds
continue to be supersymmetric, despite the fact that $K_7$ is deformed
away from $G_2$ holonomy, precisely because the supercovariant
derivative appearing in the gravitino transformation rule acquires the
correction appearing in (\ref{covmod}).

   At the leading order, the embedding of the $G_2$ holonomy group in
$SO(7)$ is such that the {\bf 8} spinor representation of $SO(7)$
decomposes to {\bf 1} + {\bf 7}, with the singlet indicating the
occurrence of a covariantly-constant spinor, \ie a Killing spinor, in
the $G_2$-holonomy background.  To study the situation once the string
or M-theory corrections are taken into account, we shall take a
specific example of a $G_2$-holonomy metric on $K_7$, and study the
generalised holonomy after the deformation of the metric implied by
the corrected equations of motion has been determined.

   As a specific example, we take the
cohomogeneity one 7-metric with $S^3\times S^3$ principal orbits,
which was first constructed in \cite{brysal,gibpagpop}:
\be
ds_7^2 = dr^2 + a^2\, (\sigma_i-\Sigma_i)^2 + b^2\, (\sigma_i+\Sigma_i)^2\,,
\ee
where $\sigma_i$ and $\Sigma_i$ are left-invariant 1-forms for two copies
of $SU(2)$.  The metric has $G_2$ holonomy if
\be
\fft{a'}{a} + \fft{b}{2a^2}=0\,,\qquad
\fft{b'}{b}- \fft{b}{4a^2} + \fft1{4b}=0\,.
\ee

    In this example, the effect of the string corrections at order $\alp$
is simply to deform the metric on $K_7$, as discussed in \cite{g2paper}.
Since no form fields are involved in the deformed background, we need
simply consider the modified covariant derivative appearing in
(\ref{covmod}).  Before including the $\alp$ corrections, a straightforward
calculation shows that the integrability condition
\be
[\nabla_i,\nabla_j] = \ft14 R_{ijk\ell}\, \Gamma^{k\ell}
\ee
selects the 14-dimensional subset $(X_a,H_1,H_2)$, $1\le a\le 12$, of the 21
$SO(7)$ generators $\Gamma_{ij}$, given by
\bea
&&X_1= \Gamma_{0\hat 1} - \Gamma_{23}\,,\qquad
X_2= \Gamma_{0\hat 2} - \Gamma_{31}\,,\qquad
X_3= \Gamma_{0\hat 3} - \Gamma_{12}\,,\nn\\
&&X_4= \Gamma_{0\hat 1} + \Gamma_{\hat2\hat3}\,,\qquad
X_5= \Gamma_{0\hat 2} + \Gamma_{\hat3\hat1}\,,\qquad
X_6= \Gamma_{0\hat 3} + \Gamma_{\hat1\hat2}\,,\nn\\
&&X_7= \Gamma_{01} - \Gamma_{2\hat3}\,,\qquad
X_8= \Gamma_{02} - \Gamma_{3\hat1}\,,\qquad
X_9= \Gamma_{03} - \Gamma_{1\hat2}\,,\nn\\
&&X_{10}= \Gamma_{01} + \Gamma_{3\hat2}\,,\qquad
X_{11}= \Gamma_{02} + \Gamma_{1\hat3}\,,\qquad
X_{12}= \Gamma_{03} + \Gamma_{2\hat1}\,,\nn\\
&& H_1 = \ft{\im}{2} (\Gamma_{1\hat1} - \Gamma_{2\hat2})\,,\qquad
H_2 = \ft{\im}{2} (\Gamma_{2\hat2} - \Gamma_{3\hat3})\,,
\eea
where we use the notation that the indices $i=(\hat1,\hat2,\hat3)$ denote
$i=(4,5,6)$, in order to emphasise the cyclic symmetry of the definitions.
A straightforward calculation shows that $(X_a,H_1,H_2)$ generate the
expected $G_2$ holonomy algebra, with the simple-root generators given by
\be
E_{\alpha_1} = X_1 - X_4 -\im\, X_7 + \im\, X_{10}\,,\qquad
E_{\alpha_2} = X_3 + X_6 -\im\, X_9 - \im\, X_{12}\,,
\ee
and having weights $\alpha_1= (1,-2)$, $\alpha_2= (0,1)$ under the Cartan
generators $(H_1,H_2)$.  Taking the Cartan-Killing metric to be
\be
g_{ij} = \ft12 \tr(H_i H_j)= \pmatrix{2 & -1\cr -1 & 2}\,,\qquad
g^{ij} = \ft13 \pmatrix{2 & 1\cr 1 & 2}\,,
\ee
we find that $\alpha_1^2= 2$, $\alpha_2^2=\ft23$ and $\alpha_1\cdot \alpha_2
=-1$.  The remaining positive-root generators are given by
\bea
&&E_{\a_1 +\a_2} = X_2 + X_5 -\im\, X_8 -\im\, X_{11}\,,\quad
E_{\a_1 +2\a_2} = X_1 + X_4 +\im\, X_7 +\im\, X_{10}\,,\\
&&E_{\a_1 +3\a_2} = X_2 - X_5 -\im\, X_8 +\im\, X_{11}\,,\quad
E_{2\a_1 +3\a_2} = X_3 - X_6 +\im\, X_8 -\im\, X_{12}\,.\nn
\eea

   At order $\alp$, we find that the correction term in the integrability
condition (\ref{int1}) produces 7 further generators $(Y_a,H_3)$,
$1\le a\le 6$, given by
\bea
&& Y_1 = -\im \, \Gamma_1 + \Gamma_{\hat2\hat3}\,,\quad
Y_2 = -\im \, \Gamma_2 + \Gamma_{\hat3\hat1}\,,\quad
Y_3 = -\im \, \Gamma_3 + \Gamma_{\hat1\hat2}\,,\nn\\
&& Y_4 = \im \, \Gamma_{\hat1} + \Gamma_{3\hat2}\,,\quad
Y_5 = \im \, \Gamma_{\hat2} + \Gamma_{1\hat3}\,,\quad
Y_6 = \im \, \Gamma_{\hat3} + \Gamma_{2\hat1}\,,\nn\\
&& H_3 = \ft12 (\Gamma_0 + \im\, \Gamma_{3\hat 3})\,.
\eea
These, together with the original $G_2$ generators $(X_a,H_1,H_2)$ that
arose at leading order, close on the algebra of $SO(7)$.  This, therefore,
is the generalised transverse holonomy group for the deformed $G_2$ metric.

    Specifically, we find that the simple roots are given by
\bea
E_{\a_1} &=& Y_3 - \im\, Y_6 + X_3 -\im\, X_9\,,\nn\\
E_{\a_2} &=& X_1 - X_4 -\im\, X_7 +\im\, X_{10}\,,\nn\\
E_{\a_3} &=& Y_3 - \im\, Y_6 - X_6 +\im\, X_{12}\,,
\eea
and that these have weights $\a_1=(0,1,0)$, $\a_2=(1,-2,1)$ and
$\a_3=(0,1,-2)$ under the Cartan generators $(H_1,H_2,H_3)$.  With the Cartan
Killing metric
\be
g_{ij} \equiv \ft12\tr(H_i H_j) = \pmatrix{2 & -1 & 0\cr
                                           -1 & 2 & -1\cr
                                           0 & -1 & 2}\,,\qquad
  g^{ij} = \ft14\pmatrix{ 3 & 2 & 1\cr
                         2 & 4 & 2\cr
                         1 & 2 & 3}\,,
\ee
we find $\a_1^2 = 1$, $\a_2^2=\a_3^2=2$, $\a_1\cdot \a_2= \a_2\cdot\a_3=-1$,
$\a_1\cdot\a_3=0$, and thus we recognise the algebra of $SO(7)$.  The
remaining positive-root generators are given by
\bea
E_{\a_1+\a_2} &=& Y_2 - \im\, Y_5 + X_2 -\im\, X_8\,,\nn\\
E_{\a_2+\a_3} &=& Y_2 - \im\, Y_5 - X_5 +\im\, X_{11}\,,\nn\\
E_{2\a_1+\a_2} &=& Y_1 + \im\, Y_4 - X_4 -\im\, X_{10}\,,\nn\\
E_{\a_1+\a_2+\a_3} &=& Y_1 + \im\, Y_4 + X_1 + \im\, X_7\,,\nn\\
E_{2\a_1+\a_2+\a_3} &=& X_2 - X_5 - \im\, X_8 +\im\, X_{11}\,,\nn\\
E_{2\a_1+2\a_2+\a_3} &=& X_3 - X_6 + \im\, X_9 -\im\, X_{12}\,.
\eea

   To summarise, we saw in section \ref{n7structuresec} that in the
seven-dimensional case the generalised transverse structure group for
purely gravitational backgrounds is $SO(8)$, enhanced from the
standard Riemannian group $SO(7)$ which occurs at leading order.  In
this section we have shown that the generalised transverse holonomy
group for a deformed $G_2$-holonomy background enlarges to $SO(7)$.
This is consistent with the continued existence of a Killing spinor,
with spinors in the {\bf 8} representation of $SO(8)$ decomposing to
{\bf 1} + {\bf 7} under the generalised transverse holonomy subgroup
$SO(7)$.

\section{Generalisation of Spin(7) Transverse Holonomy}

   In section \ref{n8structuresec}, we showed that the
structure group in the eight-dimensional transverse space for backgrounds
of the form (Minkowski)$\times K_8$ is augmented from
the usual $SO(8)$ Riemannian group to $SO(8)_+\times SO(8)_-$ when the
effects of higher-order corrections in string or M-theory are taken into
account.  In this section, we shall study the corresponding generalised
transverse holonomy group that arises when one studies the corrections to
leading-order supersymmetric Spin(7) holonomy backgrounds.

   For a concrete example, we shall start from the Spin(7) holonomy 
metric on the $\R^4$ bundle over $S^4$, which was first constructed in
\cite{brysal,gibpagpop}.  This metric is contained within the 
cohomogeneity-1 class of metrics
\be
ds_8^2 = dr^2 + a(r)^2 \, R_i^2 + b(r)^2\, P_a^2\,,
\ee
where $R_i$, $1\le i\le 3$ and $P_a$, $1\le a \le 4$, are left-invariant
1-forms in the coset $S^7\sim SO(5)/SO(3)$, as described in \cite{cvgilupo}.
It is convenient to use an orthonormal basis
\be
e^0=dr\,,\quad e^i = a\, R_i\,,\quad\, e^4= b\, P_0\qquad
   e^{\hat i} = b\, P_{\hat i}\,,
\ee
where $\hat i$ corresponds to the index values $5,6,7$ as $i$ ranges
over $1,2,3$.  At leading order, prior to the inclusion of
higher-order corrections, the metric has Riemannian holonomy Spin(7)
if the functions $a$ and $b$ satisfy the first-order equations \cite{cvgilupo}
\be
a'= 1 - \fft{a^2}{2b^2}\,,\qquad b' = \fft{3a}{4b} \,.
\ee
After some algebra, we find that the commutators $[\nabla_i,\nabla_j] = \ft14
R_{ijk\ell}\, \Gamma^{k\ell}$ indeed select a 21-dimensional subset of the
28 $SO(8)$ generators $\Gamma_{ij}$, which generate the algebra of Spin(7)
$\subset SO(8)$. These are given by the 6 generators
\bea
&& \Gamma_{01} + \Gamma_{4 \hat 1}\,,\qquad
\Gamma_{23} + \Gamma_{4 \hat 1}\,,\qquad
\Gamma_{4 \hat 1} - \Gamma_{\hat 2 \hat 3}\,,\nn\\
&& \Gamma_{0 \hat 1} - \Gamma_{3 \hat 2}\,,\qquad
  \Gamma_{0\hat 1} + \Gamma_{2\hat 3}\,,\qquad
    2\Gamma_{14}- \Gamma_{2\hat 3} + \Gamma_{3\hat 2}\,,
\eea
together with 12 more obtained by cyclic permutation of $1,2,3$, and a
3 further generators given by
\be
H_i = \ft{\im}{2} (\Gamma_{04} - \Gamma_{i\, \hat i})\,, \qquad
1=1,2,3\,.
\ee
We can take the $H_i$ to be the Cartan generators, and we find that the
remaining 18 generators can be organised into 9 positive-root generators
and 9 negative-roots generators, with the three simple-root generators
\bea
E_{\a_1} &=& -\im\, \Gamma_{01} + \im\, \Gamma_{23} +
\im\, \Gamma_{4 \hat 1} - \im\, \Gamma_{\hat 2\hat3} - \Gamma_{0\hat 1}
+ \Gamma_{14} - \Gamma_{2\hat3} + \Gamma_{3\hat 2}\,,\nn\\
E_{\a_2} &=& \im\, \Gamma_{12} + \im\, \Gamma_{\hat1\hat2} + \Gamma_{1\hat 2}
   + \Gamma_{2\hat 1}\,,\\
E_{\a_3} &=& \im\, \Gamma_{23} + \im\, \Gamma_{\hat2\hat3}
   + \Gamma_{2\hat3} + \Gamma_{3\hat2}\,.\nn
\eea
The remaining roots
\be
\a_1+\a_2\,,\quad \a_2+\a_3\,,\quad
2\a_1+\a_2\,,\quad \a_1+\a_2+\a_3\,,\quad
2\a_1+\a+2+\a_3\,,\quad
2\a_1+2\a_2 +\a_3\,,
\ee
arising from the corresponding commutation of the simple roots, complete the
positive roots of Spin(7) and, together with their conjugates, fill out the
generators found above.

    When the string or M-theory corrections are taken into account, we
obtain extra terms arising from the commutators of the supercovariant
derivatives $\hat D_i$ given in (\ref{did8}).  These terms have a
rather specific structure, which can be described as follows.
Regardless of whether we consider tree-level string corrections, or
one-loop string/M-theory corrections, the extra terms coming from
$\nabla_i$ itself in the deformed background combine with the terms from
$Q_i$ in such a way that we get precisely the chirally-projected
matrices
\be
\ft12(1+\Gamma_9)\,\Gamma_{ij}\,.\label{chiralp}
\ee
These generate the full $SO(8)_+$ algebra.  In the case where we
consider one-loop string or M-theory corrections, the functions $f$ and
$A$ are non-zero, but related to each other according to $f=3A$ 
\cite{spin7paper}.
This means that again, the further terms arising in the corrected
commutators all have the chiral structure (\ref{chiralp}).  The crucial
point is that as far as the negative chiral projection
$\ft12(1-\Gamma_9)\, \Gamma_{ij}$ is concerned, no new contributions arise
when the higher-order corrections are taken into account.

   The upshot of the above discussion is that whether one looks at the
tree-level corrections in string theory, or at the one-loop string
or M-theory corrections, the effect is to introduce the full complement of
28 generators of $SO(8)_+$, but to leave the structure of negative-chiral
projected generators $\ft12(1-\Gamma_9)\, \Gamma_{ij}$ unaltered from their
leading-order uncorrected form.  In other words, the full set of
Dirac matrices that arises for the corrected solutions comprises the
28 generators of  $SO(8)_+$, and the 21 generators of Spin(7)$_-$.
In other words, the generalised transverse holonomy group for the deformed
(Minkowski)$\times K_8$ backgrounds is the augmentation of the original
undeformed Spin(7)$\subset SO(8)$ to
\be
SO(8)_+ \times \hbox{Spin(7)}_- \subset
    SO(8)_+ \times SO(8)_-\,.
\ee
Note that again we have the property that there is a singlet in the
decomposition of spinors of the generalised structure group under the
generalised holonomy subgroup, which is consistent with the findings
of \cite{spin7paper} that supersymmetry is preserved by the
higher-order corrections.

\section{Generalisation of $SU(5)$ Transverse Holonomy}

  We now turn to the discussion of the generalised transverse holonomy
group for deformations of (Minkowski)$_1\times K_{10}$ supersymmetric
backgrounds in M-theory, where $K_{10}$ is initially a Ricci-flat
K\"ahler 10-manifold, with $SU(5)$ holonomy.  The details of the 
corrected solutions were discussed in \cite{spin7paper}.  It was found that
again, as in the lower-dimensional cases, the deformed solution remains
supersymmetric, even though the 10-dimensional metric ceases to have 
$SU(5)$ special holonomy.  In fact in this case, the deformed metric
does not even remain K\"ahler, although $K_{10}$ is still a complex
manifold \cite{spin7paper}.  It should be noted that the deformed
eleven-dimensional solution takes the warped-product form described by
(\ref{11to10}), and furthermore, the 4-form field strength is
non-vanishing, taking the form $\hat F_\4 = G_\3\wedge dt$ \cite{spin7paper}.

   For a concrete example, we shall consider a ten-dimensional metric
of the form
\be
ds_{10}^2 = dr^2 + a^2\, (d\Omega_1^2 + d\Omega_2^2
  + d\Omega_3^2 + d\Omega_4^2) + b^2\, (d\tau + {\cal A})^2\,,\label{su5met}
\ee
where $d\Omega_i^2$ denotes the metric on the $i$'th of four unit 2-spheres,
and $d{\cal A}= \sum_i \Omega_i$, where $\Omega_i$ is the volume form on the
$i$'th 2-sphere.  The functions $a$, $b$ and $c$ depend only on $r$, and 
the metric has $SU(5)$ holonomy if they satisfy the first-order equations
\be
a'= \fft{b}{2a}\,,\qquad b'= 1 - \fft{2b^2}{a^2}\,.\label{su5fo}
\ee
In what follows, we shall adopt a vielbein basis given by
\bea
&&e^1= a\, d\theta_1\,, \qquad e^2= a\, \sin\theta_1\, d\phi_1\,,\qquad
e^3= a\, d\theta_2\,, \qquad e^4= a\, \sin\theta_2\, d\phi_2\,,\nn\\
&&e^5= a\, d\theta_3\,, \qquad e^6= a\, \sin\theta_3\, d\phi_3\,,\qquad
e^7= a\, d\theta_4\,, \qquad e^8= a\, \sin\theta_4\, d\phi_4\,,\nn\\
&&e^0=b\, (d\tau + {\cal A})\,,\qquad e^9= dr\,.
\eea

    It is straightforward (with the aid of a computer) to work out the
corrected first-order equations that arise from imposing the
conditions derived in \cite{spin7paper}, and to determine the warp
factor $A$ in (\ref{11to10}) and the non-vanishing 4-form field
strength for this deformed solution.  From this information, one can
then construct the deformed supercovariant derivative operator, and
hence read off the generators of the generalised holonomy group.  

   We find that after including multiple commutators until closure is
achieved, there are in total 217 generators of the generalised
transverse holonomy group in our example.  They are all, of course,
contained within the set of 510 generators of the $SL(16,\C)$
generalised transverse structure group listed in (\ref{sl16c}).
Before presenting some details of our results, we shall first
summarise the conclusions.  After manipulations that are again best
performed with the aid of a computer, we find that there is a Cartan
subalgebra of dimension 17, and there are a further 120 of the 217
generators that all commute with each other.  The remaining 80
generators, together with 16 of the Cartan generators, give rise to
the algebra $SL(5,\C)\times SL(5,\C)$. The 17'th Cartan generator is
compact, and we find that the complete group is of the general form
\be
[U(1)\times SL(5,\C)\times SL(5,\C) ] \ltimes \R^{120}\,,\label{217group}
\ee
where the symbol $\ltimes$ denotes a semi-direct product.
Specifically, we find that
the 120 mutually-commuting factors in $\R^{120}$ assemble in the form
\be
\C^{(10,1)}_{1} \oplus \C^{(10,5)}_3\,,
\ee
where the superscripts denote the representations under the left and right
$SL(5,\C)$ factors, and the subscripts denote the charges under the 17'th
Cartan generator associated with the $U(1)$ factor in (\ref{217group}).

   The construction of the 16 Cartan generators of $SL(5,\C)\times 
SL(5,\C)$ is as follows.  Defining
\bea
&& h_1= \Gamma_{12} + \Gamma_{09}\,,\quad
h_2= \Gamma_{34} + \Gamma_{09}\,,\quad
h_3= \Gamma_{56} + \Gamma_{09}\,,\quad
h_4= \Gamma_{78} + \Gamma_{09}\,,\nn\\
&& h_5 = \Gamma_{12}\, (3\Gamma_{09} - \Gamma_{34} -\Gamma_{56} -
             \Gamma_{78}) + 2(\Gamma_{3456} + \Gamma_{3478} +
       \Gamma_{5678})\,,\nn\\
&& h_6 = \Gamma_{34}\, (3\Gamma_{09} - \Gamma_{12} -\Gamma_{56} -
             \Gamma_{78}) + 2(\Gamma_{1256} + \Gamma_{1278} +
       \Gamma_{5678})\,,\nn\\
&& h_7 = \Gamma_{56}\, (3\Gamma_{09} - \Gamma_{12} -\Gamma_{34} -
             \Gamma_{78}) + 2(\Gamma_{1234} + \Gamma_{1278} +
       \Gamma_{3478})\,,\nn\\
&& h_8 = \Gamma_{78}\, (3\Gamma_{09} - \Gamma_{12} -\Gamma_{34} -
             \Gamma_{56}) + 2(\Gamma_{1234} + \Gamma_{1256} +
       \Gamma_{3456})\,,
\eea
we find that the Cartan generators of the left-hand $SL(5,\C)$ can be taken
to be
\bea
H_1 &=& \ft{3\im}{8}\, h_1 - \ft1{24} (h_6+h_7+h_8)\Gamma_{11}\,,\nn\\
H_2 &=& \ft{3\im}{8}\, h_2 - \ft1{24} (h_5+h_7+h_8)\Gamma_{11}\,,\nn\\
H_3 &=& \ft{3\im}{8}\, h_3 - \ft1{24} (h_5+h_6+h_8)\Gamma_{11}\,,\nn\\
H_4 &=& \ft{3\im}{8}\, h_4 - \ft1{24} (h_5+h_6+h_7)\Gamma_{11}\,.
\eea
The generators of the right-hand $SL(5,\C)$ factor can be taken to 
be
\bea
\wtd H_1 &=& -\ft{\im}{8}\, h_1 + \ft1{24} (h_6+h_7+h_8)\Gamma_{11}\,,\nn\\
\wtd H_2 &=& -\ft{\im}{8}\, h_2 + \ft1{24} (h_5+h_7+h_8)\Gamma_{11}\,,\nn\\
\wtd H_3 &=& -\ft{\im}{8}\, h_3 + \ft1{24} (h_5+h_6+h_8)\Gamma_{11}\,,\nn\\
\wtd H_4 &=& -\ft{\im}{8}\, h_4 + \ft1{24} (h_5+h_6+h_7)\Gamma_{11}\,.
\eea
The 17'th Cartan generator, associated with the $U(1)$ factor in
(\ref{217group}), is given by
\bea
H_{17} &=& \ft{\im}2\, \Gamma_{09} - \ft14(2 \im +
\Gamma_{09}\Gamma_{11} )\,
   (\Gamma_{12} + \Gamma_{34} +\Gamma_{56} + \Gamma_{78})\nn\\
&&+ \ft14 (\Gamma_{1234} + \Gamma_{1256} + \Gamma_{1278} + \Gamma_{3456} 
   + \Gamma_{3478} + \Gamma_{5678})\Gamma_{11}\,.
\eea

   We shall not present all the remaining $SL(5,\C)\times SL(5,\C)$
generators, but just those corresponding to the simple roots, from which the
rest follow by commutation.  For the left-hand $SL(5,\C)$ we have the
simple-root generators
\bea
E_1 &=& (\Gamma_{14} + \im\, \Gamma_{13} - \Gamma_{23} +\im\, \Gamma_{24})
 [3 + \im\, (\Gamma_{09} - \Gamma_{56} - \Gamma_{78})\Gamma_{11}]\,,\nn\\
E_2 &=& (\Gamma_{36} + \im\, \Gamma_{35} - \Gamma_{45} +\im\, \Gamma_{46})
 [3 + \im\, (\Gamma_{09} - \Gamma_{12} - \Gamma_{78})\Gamma_{11}]\,,\nn\\
E_3 &=& (\Gamma_{58} + \im\, \Gamma_{57} - \Gamma_{67} +\im\, \Gamma_{68})
 [3 + \im\, (\Gamma_{09} - \Gamma_{12} - \Gamma_{34})\Gamma_{11}]\,,\nn\\
E_4 &=& (\Gamma_{07} + \im\, \Gamma_{08} + \Gamma_{89} -\im\, \Gamma_{79})
 [-3 + \im\, (\Gamma_{12} +\Gamma_{34} + \Gamma_{56})\Gamma_{11}]\,.
\eea
For the right-hand $SL(5,\C)$, we have the simple-root generators
\bea
\wtd E_1 &=& (\Gamma_{14} + \im\, \Gamma_{13} - \Gamma_{23} +\im\, \Gamma_{24})
 [-1 + \im\, (\Gamma_{09} - \Gamma_{56} - \Gamma_{78})\Gamma_{11}]\,,\nn\\
\wtd E_2 &=& (\Gamma_{36} + \im\, \Gamma_{35} - \Gamma_{45} +\im\, \Gamma_{46})
 [-1 + \im\, (\Gamma_{09} - \Gamma_{12} - \Gamma_{78})\Gamma_{11}]\,,\nn\\
\wtd E_3 &=& (\Gamma_{58} + \im\, \Gamma_{57} - \Gamma_{67} +\im\, \Gamma_{68})
 [-1 + \im\, (\Gamma_{09} - \Gamma_{12} - \Gamma_{34})\Gamma_{11}]\,,\nn\\
\wtd E_4 &=& (\Gamma_{07} + \im\, \Gamma_{08} + \Gamma_{89} -\im\, \Gamma_{79})
 [1 + \im\, (\Gamma_{12} +\Gamma_{34} + \Gamma_{56})\Gamma_{11}]\,.
\eea
The simple root vectors are given by $\alpha_1=\td \alpha_1=\{1,-1,0,0\}$,
$\alpha_2=\td \alpha_2=\{0, 1,-1,0\}$, $\alpha_3=\td \alpha_3=\{0,0,1,-1\}$
and $\alpha_4=\td \alpha_4=\{1,1,1,2\}$ respectively.
The Cartan Killing metric for the left-hand $SL(5,\C)$ is given by
\be
g_{ij}=\ft12\tr(H_iH_j) =3\pmatrix{2 & 1 & 1 & 1\cr
                                   1 & 2 & 1 & 1\cr
                                   1 & 1 & 2 & 1\cr
                                   1 & 1 & 1 & 2}\,,\qquad
g^{ij}=\ft1{15} \pmatrix{ 4 & -1 & -1 & -1 \cr
                   -1 &  4 & -1 & -1 \cr
                   -1 & -1 &  4 & -1 \cr
                   -1 & -1 & -1 &  4}\,.
\ee
The Cartan Killing metric for the right-hand $SL(5,\C)$ is given by
$\td g_{ij}=\ft13 g_{ij}$ and $\td g^{ij}=3 g^{ij}$.

    Finally, we present the $(10,1)$ and $(10,5)$ of 
mutually-commuting $\C$ factors.
The $(10,1)$ representation can be generated from the
highest-weight generator, whose weight-vector is $\{1,1,1,0\}$; it is 
given by
\bea
V&=&(1-\Gamma_{10}) (x + \im\, y)\,,\nn\\
x &=& \Gamma_{08} + \Gamma_{79} - (\Gamma_{07} - \Gamma_{89})
(\Gamma_{12} + \Gamma_{34} + \Gamma_{56})\,,\nn\\
y&=& \Gamma_{07} - \Gamma_{89} + (\Gamma_{08} + \Gamma_{79})
(\Gamma_{12} + \Gamma_{34} + \Gamma_{56})\,.
\eea
It is straightforward to verify that the 10 generators commute with
the right-hand $SL(5,\C)$.  These generators all have eigenvalue $+1$ under
$H_{17}$, whilst their conjugates $(1+G_{10}) (x - \im y)$
have eigenvalue $-1$.

    The $(10,5)$ generators can be generated analogously from the 
highest-weight generator
\bea
U &=& (1 - \Gamma_{10})\Big( (\Gamma_{01} + \Gamma_{29})(\Gamma_{36} +
\Gamma_{45}) + (\Gamma_{02} -\Gamma_{19})(\Gamma_{35}-\Gamma_{46})
\nn\\
&&\qquad\qquad+
\im\, (\Gamma_{01} + \Gamma_{29})(-\Gamma_{35} + \Gamma_{46}) + \im\, 
(\Gamma_{02} - \Gamma_{19}) (\Gamma_{36} + \Gamma_{45})\Big)\,,
\eea
which, has weight-vector $\{1,1,1,0\}$ under the left-hand $SL(5,\C)$
and $\{1,1,1,1\}$ under the right-hand $SL(5,\C)$.  The resulting
$(10,5)$ generators all have weight 3 under the Cartan generator $H_{17}$.
This completes the demonstration that for the supersymmetric 
(Minkowski)$\times K_{10}$ background that we have considered here, the
generalised transverse holonomy group is
\be
[U(1)\times SL(5,\C)\times SL(5,\C) ] \ltimes [\C_1^{(10,1)} \oplus
             \C_3^{(10,5)}]\,.\label{217groupc}
\ee

   Note that just as the transverse structure group
$SL(16,\C)$ of this (Minkowski)$\times K_{10}$
background is non-compact, so too is the generalised transverse 
holonomy group that we have
obtained here.  Although it is perhaps not immediately
apparent, there is, as the preservation of supersymmetry tells us
there must be, a singlet in the decomposition of the 16-dimensional
representation of the generalised transverse structure group
$SL(16,\C)$ under the generalised holonomy subgroup (\ref{217groupc}).
An easy, if rather mechanical, way to see this is simply to note that
there exists a spinor which is annihilated by all 217 of the
generalised holonomy-group generators.

   The result that we have obtained here for the generalised
transverse holonomy group (\ref{217groupc}) is considerably more
complicated, and at the same time more degenerate, than the type of
groups we encountered for lower-dimensional transverse spaces.  One
might suspect that this could be a consequence of having chosen in
(\ref{su5met}) a rather special and simple example for the type of
deformed $SU(5)$-holonomy metric.  It might well be that with a more
generic type of example, for which the curvature tensor had a larger
number of non-zero eigenvalues, one would find a less
degenerate-looking generalised holonomy group.  Unfortunately the
computational difficulties associated with using a more complicated
example have prevented us from pursuing this question further.

\section{Conclusions}

    In this paper, we have studied the effect of higher-order
corrections in string and M-theory on the holonomy groups of
supersymmetric backgrounds.  In order to do this, we have made use of
the notion of generalised holonomy, which was introduced in
\cite{dust,duli1,hull}.  In the earlier works on the subject, the
focus was on supersymmetric backgrounds in string or M-theory at the
leading order, but where form fields were turned on, leading to the
need to generalise the usual notion of Riemannian structure and
holonomy groups.  In the present paper, by contrast, our principal
focus has been on purely gravitational backgrounds, but taking into
account the effect of higher-order corrections to the equations of
motion, and the supersymmetry transformation rules. The latter imply
that the structure group and holonomy group will already be
generalised and enlarged, because the corrections to the supersymmetry
transformation rules introduce new gamma-matrix terms in the
supercovariant derivative.  In some cases, for transverse spaces of
dimension eight or ten, the corrected equations of motion require that
the form fields that started out zero in the leading-order solution
must become non-zero, and this can further enlarge the generalised
structure and holonomy groups.  One of the motivations for our
investigation was to study how, at the level of the generalised 
structure group and its generalised holonomy subgroup, the deformation
implied by the corrections to the string or M-theory equations can 
still yield a supersymmetric background even though the ordinary 
holonomy group is enlarged from special holonomy to generic holonomy.
Our study focused on the generalised structure groups and holonomy 
groups restricted to the transverse space, since this captures the 
essence of the non-trivial nature of the background solutions.

   Our results can be conveniently summarised in two tables.  First,
in Table 1, for completeness, we list the standard structure groups
and holonomy groups in the transverse spaces $K_n$ for dimensions
$n=6$, 7, 8 and 10.  Then, in Table 2, we list our findings for the
analogous generalised structure groups and holonomy groups for each of
the transverse dimensions $n=6$, 7, 8 and 10, where the effect of the
known string or M-theory corrections to the leading-order equations
are taken into account.  In each table, the structure group can be
thought of as the holonomy group for a generic non-supersymmetric
background, whilst the holonomy groups that we list are for
supersymmetric backgrounds.

   \bigskip\bigskip
\centerline{
\begin{tabular}{|c|c|c|} \hline
$n$ & Structure Group & Holonomy Group \\ \hline \hline
 6 & $SO(6)$ & $SU(3)$ \\  \hline
 7 & $SO(7)$ & $G_2$ \\ \hline
 8 & $SO(8)$ & Spin(7) \\ \hline
10 & $SO(10)$ & $SU(5)$ \\ \hline
\end{tabular}}
\bigskip

\noindent{\bf Table 1:} The structure group for the transverse space $K_n$
and the holonomy group for a supersymmetric background, at leading order
in string or M-theory.
\bigskip\bigskip

   \bigskip\bigskip
\centerline{
\begin{tabular}{|c|c|c|} \hline
$n$ & Structure Group & Holonomy Group \\ \hline \hline
 6 & $SO(6)\times U(1)$ & $SU(3)\times U(1)$ \\  \hline
 7 & $SO(8)$ & $SO(7)$ \\ \hline
 8 & $SO(8)_+\times SO(8)_- $ & $SO(8)_+ \times $Spin(7)$_-$ \\ \hline
10 & $SL(16,\C)$ & $[U(1)\times SL(5,\C)\times SL(5,\C)]\ltimes 
          [\C^{(10,1)}\oplus \C^{(10,5)}]$ \\ \hline
\end{tabular}}
\bigskip

\noindent{\bf Table 2:} The generalised structure group for the transverse 
space $K_n$
and the generalised holonomy group for a supersymmetric background, 
including higher-order corrections 
in string or M-theory.
\bigskip\bigskip

\section*{Acknowledgments}

    We are grateful to Paul Townsend for discussions.  C.N.P. and 
K.S.S thank the Theory Group at CERN, and K.S.S. thanks the
George P. \& Cynthia W. Mitchell Institute for Fundamental Physics, for
hospitality during the course of this work.

\end{document}